\documentstyle[twoside,11pt]{article}
\normalsize
\begin{document}
\bibliographystyle{plain}

\def\greaterthansquiggle{\raise.3ex\hbox{$>$\kern-.75em\lower1ex\hbox{$\sim$}}}
\def\lessthansquiggle{\raise.3ex\hbox{$<$\kern-.75em\lower1ex\hbox{$\sim$}}}
\newcommand{\bdi}{\begin{displaymath}}
\newcommand{\edi}{\end{displaymath}}
\newcommand{\bfi}{\begin{figure}}
\newcommand{\efi}{\end{figure}}
\newcommand{\npb}{\nopagebreak[3]}
\newcommand{\beq}{\begin{equation}}
\newcommand{\eeq}{\end{equation}}
\newcommand{\gaa}{\gamma_{\alpha}}
\newcommand{\gab}{\gamma_{\beta}}
\newcommand{\gam}{\gamma_{\mu}}
\newcommand{\gan}{\gamma_{\nu}}
\newcommand{\gaM}{\gamma^{\mu}}
\newcommand{\gaN}{\gamma^{\nu}}
\newcommand{\gaf}{\gamma_{5}}
\newcommand{\tr}{\mbox{tr}}
\newcommand{\Tr}{\mbox{\it\bf Tr}}
\newcommand{\im}{\mbox{Im}}
\newcommand{\intm}{\int_{4m^2}^{\infty}}
\newcommand{\fpi}{\frac{1}{\pi}}
\newcommand{\range}{\mbox{range}}
\newcommand{\coker}{\mbox{coker}}
\newcommand{\ch}{\mbox{ch}}
\newcommand{\beqa}{\begin{eqnarray}}
\newcommand{\eeqa}{\end{eqnarray}}
\newcommand{\no}{\nonumber}
\newcommand{\bob}{\hspace{0.2em}\rule{0.5em}{0.06em}\rule{0.06em}
{0.5em}\hspace{0.2em}}
\newcommand{\grts}{\greaterthansquiggle}
\newcommand{\lets}{\lessthansquiggle}
\def\dddot{\raisebox{1.2ex}{$\textstyle .\hspace{-.12ex}.
\hspace{-.12ex}.$}\hspace{-1.5ex}}
\def\Dddot{\raisebox{1.8ex}{$\textstyle .\hspace{-.12ex}.
\hspace{-.12ex}.$}\hspace{-1.8ex}}
\newcommand{\Un}{\underline}
\newcommand{\ol}{\overline}
\newcommand{\ra}{\rightarrow}
\newcommand{\Ra}{\Rightarrow}
\newcommand{\ve}{\varepsilon}
\newcommand{\vp}{\varphi}
\newcommand{\vt}{\vartheta}
\newcommand{\dg}{\dagger}
\newcommand{\wt}{\widetilde}
\newcommand{\wh}{\widehat}
\newcommand{\br}{\breve}
\newcommand{\A}{{\cal A}}
\newcommand{\B}{{\cal B}}
\newcommand{\C}{{\cal C}}
\newcommand{\D}{{\cal D}}
\newcommand{\F}{{\cal F}}
\newcommand{\G}{{\cal G}}
\newcommand{\Ha}{{\cal H}}
\newcommand{\K}{{\cal K}}
\newcommand{\cL}{{\cal L}}
\newcommand{\M}{{\cal M}}
\newcommand{\R}{{\cal R}}
\newcommand{\dfrac}{\displaystyle \frac}
\newcommand{\hy}{${\cal H}\! \! \! \! \circ $}
\newcommand{\h}[2]{#1\dotfill\ #2\\}
\newcommand{\tab}[3]{\parbox{2cm}{#1} #2 \dotfill\ #3\\}
\def\au{{\setbox0=\hbox{\lower1.36775ex%
\hbox{''}\kern-.05em}\dp0=.36775ex\hskip0pt\box0}}
\def\ao{{}\kern-.10em\hbox{``}}
\def\lint{\int\limits}

%
%
%
%
%
\newcommand{\dsla}{\partial\hspace{-6pt} /  }
\newcommand{\Asla}{A\hspace{-6.5pt}  /  }
\newcommand{\Dsla}{D\hspace{-7.3pt}  /  }
\newcommand{\Qsla}{Q\hspace{-7.2pt}  /  }
\newcommand{\psla}{p\hspace{-5.375pt} /   }
\newcommand{\ksla}{k\hspace{-6pt} /  }
\newcommand{\qsla}{q\hspace{-6pt} /   }
\newcommand{\asla}{a\hspace{-6.25pt} /   }
\newcommand{\eps}{\epsilon\hspace{-5.325pt} / }
\newcommand{\zsla}{z\hspace{-6pt} /  }
\newcommand{\DDsla}{D\hspace{-5.83pt}  /  }
\newcommand{\ddsla}{\partial\hspace{-4.6pt} /  }
\newcommand{\DDsq}{D\hspace{-5.83pt}  / \hspace{2.2pt} ^2 }
\newcommand{\Dsq}{D\hspace{-7.3pt}  /  \hspace{2.5pt} ^2 }
\newcommand{\AAsla}{A\hspace{-5pt}  /  }
\newcommand{\QQsla}{Q\hspace{-5.8pt}  /  }

\begin{titlepage}
\begin{flushright}
UWThPh-1994-39\\
15 Dec 1994
\end{flushright}
\vspace{2cm}
\begin{center}
{\Large \bf The Dyson-Schwinger equation for a model with
instantons -- the Schwinger model*}\\[1cm]
C. Adam \\
Institut f\"ur Theoretische Physik \\
Universit\"at Wien
\vfill
{\bf Abstract:} \\
\end{center}
Using the exact path integral solution of the Schwinger model --
a model where instantons are present --
the Dyson-Schwinger equation is shown to hold by explicit
computation. It turns out that the Dyson-Schwinger equation
separately holds for every instanton sector. This is due to
$\Theta $ invariance of the Schwinger model. \\[1cm]
PACS--No 11.10 Mn \ldots Field theory, Schwinger source theory \\
PACS--No 11.15 Tk \ldots Gauge field theories, other
nonperturbative techniques
\vfill
\noindent *) Supported by ``Fonds zur F\"orderung der wissenschaftlichen
Forschung in \"Osterreich'', Projekt Nr. P8444-TEC
\end{titlepage}

\section*{Introduction}

The Schwinger model (SM), which is ${\rm QED_2}$ with one massless
fermion, is known to be exactly soluble (\cite{Sc1}), and
explicit solutions within the operator formalism (\cite{LS1})
and the path integral (PI) approach (\cite{Jay}, \cite{SW1},
\cite{DSW}, \cite{Sm1}, \cite{Adam}, \cite{Diss}) are wellknown.
Further instantons and an instanton vacuum have to be present in
the SM in order to obtain a consistent quantization
(\cite{Adam}, \cite{Diss}). So the question of whether there is a
modification of the Dyson-Schwinger (DS) equation (\cite{IZ1}, \cite{Dy1},
\cite{Sc2}) when an instanton
vacuum exists may be answered by explicit calculation in the SM.
On one hand we will compute that for the trivial sector
(instanton number $k=0$) the DS equation holds as it must be. On
the other hand it will turn out that the DS equation even holds
for the nontrivial sector $k \not= 0$. More precisely it holds
separately for every instanton sector $k$. So, at least for the
model at hand, the DS equation may be extended to the case where
the true vacuum is not the perturbative one but an instanton or
$\Theta$ vacuum. The reason why every instanton sector separately
fulfills the DS equation is related to the $\Theta$ invariance
of the SM, as we will discuss below.

Our computations are done in flat Euclidean space $E^2$, and we
use conventions like in \cite{Adam}, \cite{Diss}, \cite{ABH}:
\bdi
g_{\mu\nu}=\delta_{\mu\nu} \quad , \quad \gaf =-i\gamma^0_{E}\gamma^1_{E}
=:-i\gamma^0 \gamma^1 \quad ,
\edi
\beq
\gam \gaf =\epsilon_{\mu\nu} \gaN \quad ,\quad
\epsilon_{\mu\nu} \epsilon^{\nu\lambda} =g_{\mu}^{\lambda}
\quad ,\quad \epsilon_{01}=-i.
\eeq
 The abelian gauge field $A_{\mu}$ is parametrized like in
\cite{Adam}, \cite{Diss},
\beq
A_{\mu}(x)=\frac{1}{e}(\partial_{\mu}\alpha (x)+\epsilon_{\mu\nu}
\partial^{\nu}\beta (x)),
\eeq
where $e$ is the dimensionfull charge. The gauge fixing $\alpha
=0$ is chosen in the sequel. $\beta$, which is the dynamical
part of $A_{\mu}$, may carry an integer instanton number
(Pontryagin index)
\beq
\nu =-\frac{i}{2\pi}\int dx\Box\beta (x)=-\frac{ie}{2\pi}\int
dx\tilde F(x) =k\in {\bf Z}.
\eeq
As representative for a "k-instanton" (wellbehaving at all
spacetime but not minimizing the action) we choose
\beq
eA_{\mu}(x)=ik\epsilon_{\mu\nu} \frac{x^{\nu}}{x^2 +\lambda^2}=
\epsilon_{\mu\nu}
\partial^{\nu}\frac{ik}{2} \ln\frac{x^2 +\lambda^2}{\lambda^2}=:
\epsilon_{\mu\nu} \partial^{\nu}\beta_k (x).
\eeq
The vacuum functional has to be summed over all $k$ instanton configurations,
\beq
Z[J_{\mu},\eta ,\bar\eta ]=\sum_{k=-\infty}^{\infty}Z_k
[J_{\mu}, \eta ,\bar\eta ],
\eeq
where
\beq
Z_k [J_{\mu},\eta ,\bar\eta ]=N\int (DA_{\mu})_k D\bar\Psi D\Psi
e^{\int dx[\bar\Psi (i\ddsla -e\AAsla )\Psi -\frac{1}{4}F_{\mu\nu}F^{\mu\nu}
+A_{\mu}J^{\mu}+\bar\Psi\eta +\bar\eta\Psi ]}.
\eeq
When one separates the zero modes of the Dirac operator and
integrates out the fermions, the resulting expression for the
vacuum functional leading to the correct quantum theory of the SM
is (see \cite{Adam}, \cite{Diss} for details)
\bdi
Z_k [J_{\mu},\eta ,\bar\eta ]=N\int D\beta \prod_{i_0 =0}^{k-1}
(\bar\eta\Psi_{i_0}^{\bar\beta})(\bar\Psi_{i_0}^{\bar\beta}\eta )\cdot
\edi
\beq
\cdot e^{i\int\bar\eta (x)G(x,y)\eta (y)dxdy}e^{\int (\frac{1}{2}\bar\beta
{\rm\bf D}\bar\beta +\frac{1}{e}\bar\beta\lambda )dx} ,
\eeq
where $\bar\beta =\beta +\beta_k $, $\beta_k $ is the
"instanton" (4) and $\beta$ has zero instanton number; further
\bdi
A_{\mu}J^{\mu}=\beta\lambda,
\edi
so
\beq
\frac{\delta}{\delta J_{\mu}(x)}= \epsilon^{\mu\nu}
\partial_{\nu}\frac{\delta}{\delta\lambda (x)};
\eeq
$G(x,y)$ is the exact fermion propagator,
\beq
G(x,y)=e^{i(\bar\beta (x)-\bar\beta (y))\gaf}G_0 (x-y),
\eeq
and $G_0$ is the free fermion propagator. The brackets mean
\beq
(\bar\eta\Psi )\equiv \int
dz\bar\eta^{\alpha}(z)\Psi^{\alpha}(z) .
\eeq
{\bf D} is the operator of the effective photon action after the
integration of the fermions (see \cite{ABH}, \cite{Diss})
\beq
{\rm\bf D}=\frac{\Box}{e^2}(\Box -\mu^2 )\quad ,\quad \mu^2 :=\frac{e^2}{\pi},
\eeq
which has the Green's function (\cite{Adam}, \cite{Diss})
\bdi
{\rm\bf G}(x-y)=\pi (D(\mu ,x-y)-D(0,x-y)),
\edi
\beq
{\rm\bf D}_{x}{\rm\bf G}(x-y)=\delta (x-y),
\eeq
$D(\mu ,x)$ and $D(0,x)$ being the massive and massless scalar
field propagators in two dimensions.

$\Psi_{i_0}^{\bar\beta}$ in (7) is a zero mode of the Dirac
operator with respect to the gauge field $\bar\beta$. It may be
related to a zero mode with respect to the "instanton"
$\beta_k$ of (4) by
\beq
\Psi_{i_0}^{\bar\beta}(x)=e^{\sigma i\beta
(x)}\Psi_{i_0}^{\beta_k}(x) ,
\eeq
where $\sigma =\pm 1$ is the eigenvalue of $\gaf $,
$\gaf \Psi_{i_0}^{\beta_k}
=\pm\Psi_{i_0}^{\beta_k}$ for $k{>\atop <}0$.
The zero modes $\Psi_{i_0}^{\beta_k}$ may be computed
(\cite{Adam}, \cite{Diss}),
\bdi
\Psi_{i_0}^{\beta_k}=\frac{1}{\sqrt{2\pi}}(x^{-})^{i_0}(\frac{x^2
+\lambda^2}{\lambda^2})^{-\frac{k}{2}}{1 \choose 0}\quad ,\quad
k>0 \quad ,\quad i_0 =0\ldots k-1 ,
\edi
\bdi
\Psi_{i_0}^{\beta_k}=\frac{1}{\sqrt{2\pi}}(x^{+})^{i_0}(\frac{x^2
+\lambda^2}{\lambda^2})^{\frac{k}{2}}{0 \choose 1}\quad ,\quad
k<0 \quad ,\quad i_0 =0\ldots \vert k\vert -1 ,
\edi
\beq
x^{+}=x_1 +ix_0 \quad ,\quad x^{-}=x_1 -ix_0 .
\eeq

\section*{The Dyson-Schwinger equations}

The general derivation of the Dyson-Schwinger (DS) equations for
QED starts with the observation that a field derivative of the
vacuum functional vanishes (\cite{IZ1}):
\beq
\frac{\delta Z[J_{\mu}, \eta ,\bar\eta ]}{\delta \bar\Psi_{\alpha}(x)}=
(\frac{\delta S}{\delta \bar\Psi_{\alpha}(x)}
[\frac{\delta}{\delta
J_{\mu}},\frac{\delta}{\delta\eta},\frac{\delta}{\delta
\bar\eta}]
+\eta_{\alpha}(x))Z[J_{\mu}, \eta ,\bar\eta ] =0
\eeq
and
\beq
\frac{\delta Z[J_{\mu}, \eta ,\bar\eta ] }{\delta A^{\nu}(x)}=0.
\eeq
Inserting the Dirac equation in (15) or the Maxwell equation in
(16), performing derivatives with respect to the sources and
setting the sources equal to zero at the end results in
equations for the exact $n$-point functions of the theory - the
DS equations.

Now the only nontrivial vacuum expectation values (VEVs)
of the SM involving instanton
contributions are VEVs of products of (pseudo)scalar currents
(\cite{Jay}, \cite{Adam}, \cite{Diss}).
The Maxwell equations however generate only gauge field and
vector current correlators, so they are not interesting for the
problem which we want to discuss - the DS equations in an
instanton vacuum. Therefore it suffices to work with equation
(15), which is the quantum analog of the Dirac equation.

The simplest nontrivial DS equation is obtained from (15) by
performing three fermion source derivatives, resulting in an
equation between fermionic two- and four-point functions. In the
explicit calculation of this DS equation we will observe a
cancellation mechanism between different Green's functions that
may be easily generalized to higher $n$-point functions. So it
is enough to deal with this DS equation (another reason for
doing the computation for the four-point function is its
importance for a Bethe-Salpeter analysis of the SM, which will
be done in a forthcoming paper). It reads
\beq
\frac{\delta}{\delta\eta_{\epsilon}(y_2)}\frac{\delta}{\delta\eta_{\delta}
(x_2)}\frac{\delta}{\delta\bar\eta_{\gamma}(y_1)}
(\eta^{\alpha}(x_1) +\gamma_{\mu}^{\alpha\beta}
(i\partial_{x_1}^{\mu}-e\frac{\delta}{\delta
J_{\mu}(x_1)})\frac{\delta}{\delta\bar\eta_{\beta}(x_1)})Z\vert_0
=0,
\eeq
where the vertical line indicates that the sources have to be
set equal to zero after the differentiation. We rewrite (17) in
three terms that we will compute separately, and they have to
sum up to zero:
\beq
[-\delta_{\delta}^{\alpha}\delta (x_2
-x_1)\frac{\delta}{\delta\eta_{\epsilon}(y_2
)}\frac{\delta}{\delta \bar\eta_{\gamma}(y_1
)}+\delta_{\epsilon}^{\alpha} \delta (y_2 -x_1)
\frac{\delta}{\delta \eta_{\delta}(x_2)}\frac{\delta}{\delta \bar\eta_{\gamma}
(y_1)}]Z\vert_0 \equiv A
\eeq
\beq
i \gamma_{\mu}^{\alpha\beta} \partial_{x_1}^{\mu}
\frac{\delta}{\delta\eta_{\epsilon}(y_2)}
\frac{\delta}{\delta\eta_{\delta}(x_2)}\frac{\delta}{\delta\bar\eta_{\gamma}
(y_1)}\frac{\delta}{\delta\bar\eta_{\beta}(x_1)}Z\vert_0 \equiv B
\eeq
\beq
-e \gamma_{\mu}^{\alpha\beta} \frac{\delta}{\delta
J_{\mu}(x_1 )}\frac{\delta}{\delta\eta_{\epsilon}(y_2)}
\frac{\delta}{\delta\eta_{\delta}(x_2)}\frac{\delta}{\delta\bar\eta_{\gamma}
(y_1)}\frac{\delta}{\delta\bar\eta_{\beta}(x_1)}Z\vert_0 \equiv C.
\eeq
Next we must remember that for a VEV of $n$ fermions and $n$
antifermions instanton sectors up to $k=\pm n$ contribute
(\cite{Jay}, \cite{Adam}, \cite{Diss}, see formula (7)).
So for the term $A$ $k=0, \pm 1$ will be important, and
for $B,C$ $k=0,\pm 1, \pm 2$.

Further the general spinor structure is important in the DS
equation only for the indices $\alpha ,\beta ,\delta $ belonging
to the $x_i$ variables, on which the DS equation acts. We
therefore may focus on the scalar component $\gamma =\epsilon$
for the $y_i$ variables without loss of information (remember
that only scalar components are important for instanton effects).
This will simplify the computations and is fixed in the sequel.

Now we are ready to start the computations. We will do them
separately for $k=0,+1,+2$ ($-1$ and $-2$ are analogous to +1, +2)
and find that the DS equation (17) separately holds for every
instanton sector.

\section*{Computation for $k=0$}

Using (7) and (8) we get for the term $C^0$ (20) (the zero
index indicates zero instanton sector)
\bdi
-e \gamma_{\mu}^{\alpha\beta} \frac{\delta}{\delta J_{\mu}(x_1 )}
\frac{\delta}{\delta\eta_{\gamma}(y_2)}
\frac{\delta}{\delta\eta_{\delta}(x_2)}\frac{\delta}{\delta\bar\eta_{\gamma}
(y_1)}\frac{\delta}{\delta\bar\eta_{\beta}(x_1)}Z\vert_0 =
\edi
\bdi
= \gamma_{\mu}^{\alpha\beta} \epsilon^{\mu\nu} N\int D\beta
(\partial_{\nu}^{x_1}\beta
(x_1))[G^{\gamma\delta}(y_1 ,x_2)G^{\beta\gamma}(x_1 ,y_2)-
\edi
\beq
-G^{\beta\delta}(x_1 ,x_2)G^{\gamma\gamma}(y_1
,y_2)]e^{\frac{1}{2} \int dz\beta{\rm\bf D}\beta},
\eeq
where the rules of Grassmann differentiation have been used.
With $G^{\gamma\gamma}=\tr G=0$ we further obtain (see (9))
\bdi
\epsilon^{\mu\nu} \gamma_{\mu}^{\alpha\beta} N\int D\beta
(\partial_{\nu}^{x_1}\beta
(x_1))[e^{i\gaf (\beta (x_1)+\beta (x_2)-\beta (y_1)-\beta
(y_2))}]^{\beta\gamma} \cdot
\edi
\beq
\cdot G_0^{\gamma\epsilon}(x_1 -y_2)G_0^{\epsilon\delta}(y_1 -x_2)
e^{\frac{1}{2}\int dz\beta{\rm\bf D}\beta}
\eeq
and, introducing the real field $\beta^{'}=i\beta$ (we omit the
prime) and using the short notation
\beq
D\mu [\beta]:=N\int D\beta e^{-\frac{1}{2}\int dz\beta{\rm\bf D}\beta}
\eeq
we get
\bdi
-i \epsilon^{\mu\nu} \gamma_{\mu}^{\alpha\beta} \int D\mu [\beta]
(\partial_{\nu}^{x_1}\beta
(x_1))\{ [{\bf 1}\cosh (\beta (x_1)+\beta (x_2)-\beta (y_1)-\beta (y_2))+
\edi
\beq
+\gaf \sinh (\beta (x_1)+\beta (x_2)-\beta (y_1)-\beta
(y_2))]G_0 (x_1 -y_2)G_0 (y_1
-x_2)\}^{\beta\delta}=
\eeq
\bdi
=-i \gamma_{\mu}^{\alpha\beta} [\gaf G_0 (x_1 -y_2)G_0
(y_1 -x_2)]^{\beta\delta} \epsilon^{\mu\nu}
\partial_{\nu}^{x_1}\int D\mu [\beta]\cdot
\edi
\beq
\cdot\cosh (\beta (x_1)+\beta (x_2)-\beta (y_1)-\beta (y_2)),
\eeq
where we used the Leibnitz rule for the $\cosh$ and the fact
that in the Gaussian PI only even powers of the variable
$\beta$ contribute. With the identity $\gam\gaf \epsilon^{\mu\nu} =-\gaN$
(see (1)) we finally arrive at
\bdi
C^0 =i \gamma_{\mu}^{\alpha\beta} G_0^{\beta\gamma}(x_1 -y_2)
G_0^{\gamma\delta}(y_1 -x_2)
\partial^{\mu}_{x_1}\int D\mu [\beta]\cdot
\edi
\beq
\cdot\cosh (\beta (x_1)+\beta (x_2)-\beta (y_1)-\beta (y_2)).
\eeq
The computation of $B^0$ is nearly identical up to formula
(22), just instead of $-\epsilon^{\mu\nu} (\partial_{\nu}^{x_1}\beta (x_1))$
we have to insert $i\partial_{x_1}^{\mu}$ acting on the whole
four-point function (see (19), (20)). Introducing again a real
$\beta$ like in (24), this time only the ${\bf 1}\cosh $ part
of the PI contributes and we finally get
\bdi
B^0 =-i \gamma_{\mu}^{\alpha\beta} \partial^{\mu}_{x_1}
G_0^{\beta\gamma}(x_1 -y_2)G_0^{\gamma\delta}(y_1 -x_2)
\int D\mu [\beta]\cdot
\edi
\beq
\cdot\cosh (\beta (x_1)+\beta (x_2)-\beta (y_1)-\beta (y_2))=
\eeq
\bdi
=-C^0 -i[\dsla_{x_1}G_0 (x_1
-y_2)]^{\alpha\gamma}G_0^{\gamma\delta}(y_1 -x_2) \int D\mu [\beta]\cdot
\edi
\beq
\cdot\cosh (\beta (x_1)+\beta (x_2)-\beta (y_1)-\beta (y_2))=
\eeq
\beq
=-C^0 -i\delta (x_1 -y_2)G_0^{\alpha\delta}(y_1 -x_2)\int D\mu [\beta]
\cosh (\beta (x_2)-\beta (y_1)).
\eeq
So the first term in $B^0$ just cancels $C^0$, whereas the
second one results in a two-point function that will cancel
$A^0$ of (18) as we now shortly compute.

Keeping $\beta$ real from the very beginning we find (because we
fixed $\gamma =\epsilon $ again the first term of (18) is zero!)
\bdi
A^0 =\delta (y_2 -x_1)\frac{\delta}{\delta\eta_{\delta}(x_2)}
\frac{\delta}{\delta\bar\eta_{\alpha}(y_1)}Z\vert_0 =
\edi
\bdi
=i\delta (y_2 -x_1)\int D\mu [\beta] G^{\alpha\delta}(y_1 -x_2)=
\edi
\beq
=i\delta (y_2 -x_1)G_0^{\alpha\delta}(y_1 -x_2)\int D\mu [\beta]\cosh
(\beta (y_1)-\beta (x_2)),
\eeq
which indeed cancels the rest of $B^0$ in (29).

So the DS equation is separately fulfilled for $k=0$, which
however is the perturbative vacuum. Therefore this result is not
too surprizing.

\section*{Computation for $k=1$}

We compute the case $k=+1$, $k=-1$ is nearly identical. In
formula (7) now one zero mode is present and we obtain for
$C^1$ (20)
\bdi
C^1 =- \epsilon^{\mu\nu} \gamma_{\mu}^{\alpha\beta}
\frac{\delta}{\delta\eta_{\gamma}(y_2)}
\frac{\delta}{\delta\eta_{\delta}(x_2)}\frac{\delta}{\delta\bar\eta_{\gamma}
(y_1)}\frac{\delta}{\delta\bar\eta_{\beta}(x_1)} N\int D\beta
(\partial_{\nu}^{x_1}\bar\beta (x_1))\cdot
\edi
\bdi
\cdot (\bar\eta\Psi_0)(\bar\Psi_0 \eta )e^{i\int dzdz^{'}
\bar\eta (z)G(z,z^{'})\eta (z^{'})}e^{\frac{1}{2}
\int dz\bar\beta{\rm\bf D}\bar\beta}=
\edi
\bdi
= -i \epsilon^{\mu\nu} \gamma_{\mu}^{\alpha\beta}
N\int D\beta (\partial_{\nu}^{x_1}\bar\beta (x_1))
[-\Psi_0^{\beta}(x_1)\bar\Psi_0^{\delta}(x_2)G^{\gamma\gamma}(y_1
,y_2)+\Psi_0^{\beta}(x_1)\bar\Psi_0^{\gamma}(y_2)G^{\gamma\delta}(y_1
,x_2)+
\edi
\beq
+\Psi_0^{\gamma}(y_1)\bar\Psi_0^{\delta}(x_2)G^{\beta\gamma}(x_1
,y_2)-\Psi_0^{\gamma }(y_1)\bar\Psi_0^{\gamma}(y_2)G^{\beta\delta}
(x_1 ,x_2)]e^{\frac{1}{2}\int dz\bar\beta{\rm\bf D}\bar\beta},
\eeq
where again $G^{\gamma\gamma}=0$ in the first term is zero.
Inserting for the $G$ formula (9), transforming the zero modes
like in (13) and introducing the zero mode projectors (see (14))
\beq
S^{\alpha\beta}(x,y):=(\Psi_0^{(\beta_1)}(x))^{\alpha}
(\bar\Psi_0^{(\beta_1)}(y))^{\beta}=\frac{1}{2\pi}P_{+}^{\alpha\beta}
e^{i(\beta_1 (x)+\beta_1 (y))}
\eeq
we get
\bdi
C^1 =-i \epsilon^{\mu\nu} \gamma_{\mu}^{\alpha\beta}
N\int D\beta (\partial_{\nu}^{x_1}\bar\beta (x_1))
[e^{i\beta (x_1)+i\beta (x_2)}S^{\beta\gamma}(x_1 ,y_2)\cdot
\edi
\bdi
\cdot (e^{i\gaf (\bar\beta (y_1)-\bar\beta (x_2))})^{\gamma\epsilon}
G_0^{\epsilon\delta}(y_1 -x_2)+e^{i\beta (y_1)+i\beta (x_2)}
S^{\gamma\delta}(y_1 ,x_2)\cdot
\edi
\bdi
\cdot
(e^{i\gaf (\bar\beta (x_1)-\bar\beta (y_2))})^{\beta\epsilon}
G_0^{\epsilon\gamma}(x_1 -y_2)-e^{i\beta (y_1)+i\beta (y_2)}
S^{\gamma\gamma}(y_1 ,y_2)\cdot
\edi
\beq
\cdot (e^{i\gaf (\bar\beta (x_1)-\bar\beta (x_2))})^{\beta\epsilon}
G_0^{\epsilon\delta}(x_1 -x_2)]e^{\frac{1}{2}\int
dz\bar\beta{\rm\bf D}\bar\beta},
\eeq
and, turning to real $\beta$ and using $\gaf P_{+}=P_{+}\gaf
=P_{+}$, we further get
\bdi
C^1 =-\frac{1}{2\pi} \epsilon^{\mu\nu} \gamma_{\mu}^{\alpha\beta}
N\int D\beta (\partial_{\nu}^{x_1}\bar\beta(x_1))
e^{-\frac{1}{2}\int dz\bar\beta{\rm\bf D}\bar\beta}\cdot
\edi
\bdi
\cdot [\sinh (\bar\beta (x_1)-\bar\beta (x_2)+\bar\beta (y_1)+\bar\beta
(y_2))P_{+}^{\beta\gamma}G_0^{\gamma\delta}(y_1 -x_2)+
\edi
\bdi
+\sinh (-\bar\beta (x_1)+\bar\beta (x_2)+\bar\beta
(y_1)+\bar\beta (y_2))P_{+}^{\gamma\delta}G_0^{\beta\gamma}(x_1 -y_2)-
\edi
\bdi
e^{\bar\beta (y_1)+\bar\beta (y_2)}\sinh (\bar\beta (x_1)-\bar\beta
(x_2))\gamma_5^{\beta\epsilon}G_0^{\epsilon\delta}(x_1 -x_2)-
\edi
\beq
-e^{\bar\beta (y_1)+\bar\beta(y_2)}\cosh (\bar\beta (x_1)-\bar\beta
(x_2))\delta^{\beta\epsilon}G_0^{\epsilon\delta}(x_1 -x_2)].
\eeq
Here the first two terms in (34) may be treated in a way
identical to that for the $k=0$ case (leading from formula (24)
to (26)). The third and fourth term, after applying Leibnitz
rule for the derivative $\partial_{\nu}^{x_1}$ and keeping track
of even $\beta$ powers only, look like $\cosh (\ldots
)\cdot\cosh (\ldots )$ and $\sinh (\ldots )\cdot \sinh (\ldots
)$ respectively. However by using the formulae
\bdi
\cosh a\cosh b=\frac{1}{2}(\cosh (a+b)+\cosh (a-b)),
\edi
\beq
\sinh a\sinh b=\frac{1}{2}(\cosh (a+b)-\cosh (a-b)),
\eeq
they may be brought into standard form. Besides, these two
terms, being proportional to {\bf 1} and $\gaf$, combine into
$P_+$, $P_-$ via formulae (35), and after some gamma matrix
algebra we find the final result
\bdi
\frac{1}{2\pi}\{\gamma_{\nu}^{\alpha\beta}
G_0^{\beta\gamma}(y_1 -x_2)P_{-}^{\gamma\delta}
\partial_{x_1}^{\nu}\int D\mu [\beta ]\cosh (\beta (x_1)-\beta
(x_2)+\beta (y_1)+\beta (y_2))+
\edi
\bdi
+\gamma_{\nu}^{\alpha\beta} G_0^{\beta\gamma}(x_1 -y_2)P_{+}^{\gamma\delta}
\partial_{x_1}^{\nu}\int D\mu [\beta ]\cosh (-\beta (x_1)+\beta
(x_2)+\beta (y_1)+\beta (y_2))-
\edi
\bdi
-\gamma_{\nu}^{\alpha\beta} G_0^{\beta\gamma}(x_1 -x_2)P_{-}^{\gamma\delta}
\partial_{x_1}^{\nu}\int D\mu [\beta ]\cosh (\beta (x_1)-\beta
(x_2)+\beta (y_1)+\beta (y_2))-
\edi
\beq
-\gamma_{\nu}^{\alpha\beta} G_0^{\beta\gamma}(x_1 -x_2)P_{+}^{\gamma\delta}
\partial_{x_1}^{\nu}\int D\mu [\beta ]\cosh (-\beta (x_1)+\beta
(x_2)+\beta (y_1)+\beta (y_2))\} .
\eeq
For the computation of $B^1$ we may again use an intermediate
result for $C^1$, (33), substitute $i\epsilon^{\mu\nu}
(\partial_{\nu}^{x_1}\bar\beta (x_1))$by $\partial_{\mu}^{x_1}$,
and find
\bdi
B^1 =\frac{1}{2\pi}\gamma_{\mu}^{\alpha\beta}
\partial_{x_1}^{\mu}\{ -P_{+}^{\beta\gamma}
G_0^{\gamma\delta}(y_1 -x_2)
\int D\mu [\beta ]\cosh (\beta (x_1)-\beta
(x_2)+\beta (y_1)+\beta (y_2))-
\edi
\bdi
-G_0^{\beta\gamma}(x_1 -y_2)P_{+}^{\gamma\delta}
\int D\mu [\beta ]\cosh (-\beta (x_1)+\beta
(x_2)+\beta (y_1)+\beta (y_2))+
\edi
\bdi
+G_0^{\beta\delta}(x_1 -x_2)\int D\mu [\beta ]\cosh (\beta (x_1)-\beta
(x_2))\cosh (\beta (y_1)+\beta (y_2))+
\edi
\beq
+\gamma_5^{\beta\gamma}G_0^{\gamma\delta}(x_1 -x_2)
\int D\mu [\beta ]\sinh (\beta (x_1)-\beta
(x_2))\sinh (\beta (y_1)+\beta (y_2))\} =
\eeq
\bdi
=-C^1 -\frac{1}{2\pi}(\dsla_{x_1}G_0 (x_1 -y_2))^{\alpha\gamma}
P_{+}^{\gamma\delta}\int D\mu [\beta ]\cosh (-\beta (x_1)+\beta
(x_2)+\beta (y_1)+\beta (y_2))+
\edi
\bdi
+(\dsla_{x_1}G_0 (x_1 -x_2))^{\alpha\delta}
\int D\mu [\beta ]\cosh (\beta (x_1)-\beta
(x_2))\cosh (\beta (y_1)+\beta (y_2))-
\edi
\beq
-(\dsla_{x_1}G_0 (x_1 -x_2))^{\alpha\gamma}\gamma_5^{\gamma\delta}
\int D\mu [\beta ]\sinh (\beta (x_1)-\beta
(x_2))\sinh (\beta (y_1)+\beta (y_2))\} =
\eeq
\bdi
=-C^1 +\frac{1}{2\pi}\{ -\delta (x_1 -y_2)P_{+}^{\alpha\delta}
\int D\mu [\beta ]\cosh (\beta
(x_2)+\beta (y_1))+
\edi
\beq
+\delta (x_1 -x_2)\delta^{\alpha\delta}
\int D\mu [\beta ]\cosh (\beta (y_1)+\beta (y_2))\} ,
\eeq
where the remaining two-point functions are again cancelled by $A^1$:
\bdi
A^1 =-\delta^{\alpha\delta}\delta (x_2 -x_1)\frac{\delta}{\delta\eta_{\gamma}
(y_2)}\frac{\delta}{\delta\bar\eta_{\gamma}(y_1)}Z\vert_0 +
\delta^{\alpha\gamma}\delta (y_2 -x_1)\frac{\delta}{\delta\eta_{\delta}(x_2)}
\frac{\delta}{\delta\bar\eta_{\gamma}(y_1)}Z\vert_0 =
\edi
\bdi
=-\delta^{\alpha\delta}\delta (x_2 -x_1)\int D\beta \Psi_0^{\gamma}(y_1)
\bar\Psi_0^{\gamma}(y_2)e^{\frac{1}{2}\int dz\bar\beta{\rm\bf D}
\bar\beta}+
\edi
\bdi
+\delta^{\alpha\gamma}\delta (y_2 -x_1)\int D\beta \Psi_0^{\gamma}(y_1)
\bar\Psi_0^{\delta}(x_2)e^{\frac{1}{2}\int dz\bar\beta{\rm\bf D}\bar\beta}=
\edi
\bdi
=\frac{1}{2\pi}\{ -\delta^{\alpha\delta}\delta (x_2 -x_1)
\int D\mu [\beta ]\cosh (\beta (y_1)+\beta (y_2))+
\edi
\beq
+P_{+}^{\alpha\delta}\delta (y_2 -x_1)
\int D\mu [\beta ]\cosh (\beta
(x_2)+\beta (y_1))\} ,
\eeq
which we wanted to prove.

The computation for $k=-1$ is analogous.

\section*{Computation for k=2}

For $k=+2$ we find for $C^2$
\bdi
C^2 =-\gamma_{\mu}^{\alpha\beta}
\epsilon^{\mu\nu}\frac{\delta}{\delta\eta_{\gamma}(y_2)}
\frac{\delta}{\delta\eta_{\delta}(x_2)}\frac{\delta}{\delta\bar\eta_{\gamma}
(y_1)}\frac{\delta}{\delta\bar\eta_{\beta}(x_1)}N\int D\beta
(\partial_{\nu}^{x_1}\bar\beta (x_1))\cdot
\edi
\bdi
\cdot (\bar\eta\Psi_0 )(\bar\Psi_0 \eta )(\bar\eta\Psi_1
)(\bar\Psi_1 \eta )e^{\frac{1}{2}\int dz\bar\beta{\rm\bf D}\bar\beta}=
\edi
\bdi
=-\gamma_{\mu}^{\alpha\beta} \epsilon^{\mu\nu}
\int D\beta (\partial_{\nu}^{x_1}\bar\beta (x_1))
e^{\frac{1}{2}\int dz\bar\beta{\rm\bf D}\bar\beta}\cdot
\edi
\bdi
\cdot [\Psi_0^{\gamma}(y_1)\bar\Psi_0^{\delta}(x_2)\Psi_1^{\beta}(x_1)
\bar\Psi_1^{\gamma}(y_2) +\Psi_0^{\beta}(x_1)\bar\Psi_0^{\gamma}(y_2)
\Psi_1^{\gamma}(y_1)\bar\Psi_1^{\delta}(x_2)-
\edi
\beq
-\Psi_0^{\gamma}(y_1)\bar\Psi_0^{\gamma}(y_2)\Psi_1^{\beta}(x_1)
\bar\Psi_1^{\delta}(x_2)-\Psi_0^{\beta}(x_1)\bar\Psi_0^{\delta}(x_2)
\Psi_1^{\gamma}(y_1)\bar\Psi_1^{\gamma}(y_2)]=
\eeq
\bdi
=-\frac{1}{4\pi^2}\gamma_{\mu}^{\alpha\beta} \epsilon^{\mu\nu}
N\int D\beta (\partial_{\nu}^{x_1}
\bar\beta (x_1))e^{\frac{1}{2}\int dz\bar\beta{\rm\bf D}\bar\beta}\cdot
\edi
\bdi
\cdot e^{i(\bar\beta (x_1)+\bar\beta (x_2)+\bar\beta
(y_1)+\bar\beta (y_2))}[P_{+}^{\gamma\delta}P_{+}^{\beta\gamma}x_1^{-}
y_2^{+}+
\edi
\beq
+P_{+}^{\beta\gamma}P_{+}^{\gamma\delta}y_1^{-}x_2^{+}-P_{+}^{\beta\delta}
x_1^{-}x_2^{+}-P_{+}^{\beta\delta}y_1^{-}y_2^{+}],
\eeq
where we used the zero modes (14) and the transformation (13).
Introducing real $\beta$ we get ($\epsilon^{\mu\nu} \gam P_{+}=-\gaN P_{+}$)
\bdi
C^2 =-\frac{i}{4\pi^2}\gamma_{\nu}^{\alpha\beta} P_{+}^{\beta\delta}
(x_1^{-}-y_1^{-})(y_2^{+}-x_2^{+})
\cdot
\edi
\beq
\cdot \partial_{x_1}^{\nu}
\int D\mu [\beta ]\cosh (\beta (x_1)+\beta (x_2)+\beta
(y_1)+\beta (y_2)).
\eeq
For the $B^2$ part we may, as usual, starting from (42), read off
\bdi
B^2 =\frac{i}{4\pi^2}\gamma_{\mu}^{\alpha\beta}
\partial_{x_1}^{\mu}P_{+}^{\beta\delta}
(x_1^{-}-y_1^{-})(y_2^{+}-x_2^{+})\cdot
\edi
\beq
\cdot \int D\mu [\beta ]\cosh (\beta (x_1)+\beta (x_2)+\beta
(y_1)+\beta (y_2)).
\eeq
Further there is no contribution from $k=2$ to the two-point
function in $A$, (18), so (43) and (44) must cancel
completely, or the derivative
\beq
\gamma_{\mu}^{\alpha\beta} \partial_{x_1}^{\mu}P_{+}^{\beta\delta}x_1^{-}
\eeq
has to vanish.
This is indeed true and is in fact used to construct the zero
modes (14) (see \cite{Adam}, \cite{Diss} for details):
\beq
\gaM \partial_{\mu}^{x}P_{+}x^{-}=
\left( \begin{array}{cc} 0 & \partial_{x^{-}} \\
\partial_{x^{+}} & 0 \end{array} \right)
\left( \begin{array}{cc} 1 & 0 \\
0 & 0 \end{array} \right) x^{-}=
\left( \begin{array}{cc} 0 & 0 \\
\partial_{x^{+}} & 0 \end{array} \right) x^{-}=0.
\eeq
So the DS equation holds for the sector $k=2$, too.

Using
\beq
x^{-}y^{+}=x_{\mu}y^{\mu}+\epsilon_{\mu\nu} x^{\mu}y^{\nu}
\eeq
we may obtain the final result (which we display for $k=\pm 2$)
\bdi
B^{\pm 2}=\frac{i}{4\pi^2}\gamma_{\mu}^{\alpha\beta}
\partial_{x_1}^{\mu}P_{\pm }^{\beta\delta}[(x_1 -y_1 )(y_2
-x_2 )\pm \epsilon_{\mu\nu} (x_1 -y_1 )^{\mu}(y_2 -x_2 )^{\nu})]\cdot
\edi
\beq
\cdot \int D\mu [\beta ]\cosh (\beta (x_1)+\beta (x_2)+\beta
(y_1)+\beta (y_2)).
\eeq

\section*{General discussion}

We showed the DS equation to hold for the simplest nontrivial
case, the fermion four-point function. For the model at hand it
is however easy to generalize the result to higher $n$-point functions.

To study the DS equation in an instanton vacuum it remains
enough to consider scalar (or pseudoscalar) bilocals of the
fermion fields, with the exception of the two fields with
variables $x_1 ,x_2 $ on which the DS equation acts. Therefore
the spinor structure reduces to simple gamma matrix products.
Further the $(\partial_{x_1}^{\nu}\beta (x_1))$ term enters the
computation of the $C$ term in a way completely analogous to our
computation, and the PI remains Gaussian.
So the partial cancellation between $B$ and $C$
persists to hold.

For a $n$-point function in the $\cosh (\cdots )$ there are $n$
$\beta (x_{i})$. Whenever two of them stem from a zero mode
projector, both $\beta $ have the same sign; whenever they stem
from a fermion propagator $G(x_{i},x_{j})$ their signs are
opposite. So when a free propagator $G_0 (x_{i},x_{j})$ is
reduced to a $\delta$-function by a derivative the corresponding
$\beta (x_{i}),\beta (x_{j})$ cancel in the $\cosh (\cdots )$
and lead to the $n-2$-point function that cancels with the term $A$.
So indeed it remains true for higher $n$-point functions that
the DS equations separately hold for every instanton sector $k$.

The reason for this separate cancellation is easy to understand.
The Schwinger model (SM) is wellknown to be independent of the
vacuum angle $\Theta $ (\cite{LS1}, \cite{Adam},
\cite{CJS}, \cite{KS1}, \cite{tH1}).
Let us remember how this feature may be seen quickly within the
PI formalism. There the $\Theta$ vacuum may be taken into
account by an additional term in the action
\beq
S\ra S+ik\Theta =S+\frac{\Theta}{2\pi}\int dx\tilde F(x)
\eeq
and by integrating over all instanton configurations (summing
over all $k$) in the vacuum functional. But the additional term
on the r.h.s. of (49) is nothing else than the index density. An
axial transformation of the fermion fields creates a similar
term in the action via the chiral anomaly:
\beq
\A =\frac{1}{\pi}\int dx\beta\tilde F.
\eeq
Usually formula (50) is derived for an infinitesimal axial
transformation $\beta$, for a constant $\beta$
formula (50) remains true for finite $\beta$. Therefore by
choosing
\beq
\beta =-\frac{\Theta}{2}
\eeq
the $\Theta$ term may be absorbed by a simple change of the
fermionic integration variables in the PI via an axial transformation.

Our instanton vacuum corresponds to the choice $\Theta =0$. For
general $\Theta $ a VEV from the sector $k$ acquires an
additional phase $e^{ik\Theta }$.
Now when the SM is invariant with respect to $\Theta $
the DS equations have to be invariant, too. When the
DS equations enforced cancellations between different instanton
sectors, those cancellation conditions would acquire a $\Theta $
dependence.
On the other hand, when the DS equations are fulfilled
separately for every instanton sector, they automatically are
$\Theta $ invariant.

Of course, the interesting question remains to be answered if
these simple features of the DS equations in an instanton vacuum
may be generalized to more complicated theories.

\end{document}